\documentclass[conference]{IEEEtran}
\IEEEoverridecommandlockouts
\usepackage{amsmath,amssymb,amsfonts}
\usepackage{algorithmic}
\usepackage{graphicx}
\usepackage{textcomp}
\usepackage{xcolor}
\usepackage{hyperref}
\hypersetup{
    colorlinks=true,
    linkcolor=black,
    filecolor=blue,      
    urlcolor=blue,
    citecolor=black,
}
\usepackage{amsmath,amssymb,amsfonts}
\usepackage{algorithmic}
\usepackage{graphicx}
\usepackage{multirow}
\usepackage{booktabs}
\usepackage{subcaption}
\usepackage{makecell}
\usepackage{pifont}
\usepackage{url}
\usepackage{soul}  
\usepackage{colortbl} 
\usepackage{threeparttable}
\usepackage[backend=biber,style=ieee,citestyle=numeric]{biblatex}
\def\BibTeX{{\rm B\kern-.05em{\sc i\kern-.025em b}\kern-.08em
    T\kern-.1667em\lower.7ex\hbox{E}\kern-.125emX}}
\begin{document}

\title{HERO: Hardware-Efficient RL-based Optimization\\ Framework for NeRF Quantization\\
\thanks{* Corresponding author: Wei Zhang. Email: $^{\dagger}$yzhangqg@connect.ust.hk, $^{*}$wei.zhang@ust.hk}
}
\vspace{-16pt}
\author{\IEEEauthorblockN{Yipu Zhang$^{1, \dagger}$, Chaofang Ma$^1$, Jinming Ge$^1$, Lin Jiang$^2$, Jiang Xu$^3$, Wei Zhang$^{1,*}$}
\IEEEauthorblockA{$^1$\textit{Department of Electronic and Computer Engineering, The Hong Kong University of Science and Technology}\\
$^2$\textit{College of Information Sciences and Engineering, Northeastern University}\\
$^3$\textit{Microelectronics Thrust, The Hong Kong University of Science and Technology (GZ)}
}
\vspace{-16pt}
}

\maketitle

\begin{abstract}
Neural Radiance Field (NeRF) has emerged as a promising 3D reconstruction method, delivering high-quality results for AR/VR applications. While quantization methods and hardware accelerators have been proposed to enhance NeRF's computational efficiency, existing approaches face crucial limitations. Current quantization methods operate without considering hardware architecture, resulting in sub-optimal solutions within the vast design space encompassing accuracy, latency, and model size. Additionally, existing NeRF accelerators heavily rely on human experts to explore this design space, making the optimization process time-consuming, inefficient, and unlikely to discover optimal solutions.
To address these challenges, we introduce HERO, a reinforcement learning framework performing hardware-aware quantization for NeRF. Our framework integrates a NeRF accelerator simulator to generate real-time hardware feedback, enabling fully automated adaptation to hardware constraints. Experimental results demonstrate that HERO achieves 1.31-1.33× better latency, 1.29-1.33× improved cost efficiency, and a more compact model size compared to CAQ, a previous state-of-the-art NeRF quantization framework. These results validate our framework's capability to effectively navigate the complex design space between hardware and algorithm requirements, discovering superior quantization policies for NeRF implementation. Code is available at \href{https://github.com/ypzhng/HERO}{https://github.com/ypzhng/HERO}.
\end{abstract}

\begin{IEEEkeywords}
Neural Radiance Field (NeRF), Hardware-software co-design, Quantization, Reinforcement learning
\end{IEEEkeywords}

\section{Introduction}
Neural Radiance Field (NeRF)~\cite{mildenhall2020nerf} is a novel 3D reconstruction method that implicitly maps input coordinates to object representations using multi-layer perceptron (MLP). While this method provides fine details for reconstruction, it demands numerous MLP computations for each input coordinate, requiring approximately 1 million FLOPs to generate only one single view~\cite{li2023instant}. 
To address this computational challenge, several works have been proposed~\cite{muller2022instant,chen2022tensorf,sun2022direct,li2023compressing}. Among these, Instant NGP~\cite{muller2022instant} has emerged as the most efficient algorithm by introducing multi-resolution hash encoding to reduce both memory consumption and computational complexity.


However, its deployment on AR/VR devices remains challenging due to two key limitations: irregular memory access patterns caused by hash encoding~\cite{li2023instant} and low computational efficiency inherent to MLP operations~\cite{liu2024content}. Recent research has addressed these challenges in neural graphics primitives through hardware-software co-design methods~\cite{li2023instant,lee2023neurex,ryu202420, zhang2025spnerf} and quantization approaches~\cite{liu2024content}. Among these methods, quantization has been proven as an effective technique to improve NeRF rendering efficiency. Specifically, quantizing the hash table reduces memory requirements and bandwidth, while quantizing MLP layers improves computational efficiency. However, current hardware accelerators typically rely on human expertise to determine quantization bit widths and often apply uniform precision across all hash table and MLP layers. This approach overlooks the potential benefits of mixed precision, which could further enhance efficiency while maintaining fine reconstruction quality. Modern computing platforms already support diverse precision formats—NVIDIA's latest GPUs accommodate FP32, FP16, FP8, FP4, INT8, Block Floating points, and Tensor cores~\cite{tensorrtSupportMatrix}, while hardware accelerators like Stripes~\cite{judd2016stripes} implement Bitserial multiplier units supporting arbitrary-precision integer multiplication. This capability presents a significant challenge for optimizing Instant NGP models, where the design space grows exponentially: with $N$ hash table levels and $L$ MLP layers, potential bit-width combinations reach $8^{(N+2L)}$ when considering options from 1 to 8 bits for hash tables, MLP activations, and weights.

\begin{table}[t]
    \centering
    \caption{Qualitative comparison table with previous work}
    \begin{threeparttable}
    \begin{tabular}{l|c c c}
    \Xhline{1px}
     Functionality & HERO (Ours) & CAQ \cite{liu2024content} & \makecell{HW accelerators \\ ~\cite{li2023instant,lee2023neurex,ryu202420,zhang2025spnerf}} \\
     \hline
     \makecell[l]{Flexible bits \\ for MLP layers}   & \ding{51} & \ding{51}  & \ding{55}  \\
     \hline
     \makecell[l]{Adjustable multiple \\ level hash table}   & \ding{51}  & \ding{55}  & \ding{55}  \\
     \hline
     \makecell[l]{Hardware feedback} & \ding{51}  & \ding{55}  & \ding{51} \\
     \hline
     \makecell[l]{Automation} & \ding{51}  & \ding{51}  & \ding{55}  \\
     \Xhline{1px}
\end{tabular}
\end{threeparttable}
    \label{tab:Compare}
    \vspace{-12pt}
\end{table}

This vast design space, encompassing both accuracy and hardware costs, presents a significant challenge for manual optimization. Current automated approaches, such as Content-Aware Quantization (CAQ)~\cite{liu2024content}, focus solely on Instant NGP algorithmic aspects without considering hardware constraints. While CAQ adaptively quantizes models based on scene-specific content and layer-wise features, it overlooks critical hardware performance metrics, making it difficult to identify suitable quantization strategies for diverse accelerator architectures.  Furthermore, existing DNN quantization frameworks~\cite{wang2019haq,kwon2024rl} only address linear layers, providing limited benefits for NeRF quantization where hash tables constitute a critical component of the overall pipeline.


Motivated by these challenges, we propose HERO, a novel automatic hardware-aware quantization framework driven by reinforcement learning (RL). We implement a custom hardware simulator based on NeuRex-style architecture~\cite{lee2023neurex} to obtain real-time latency measurements as direct hardware feedback. Our RL agent optimizes mixed-precision quantization strategies by incorporating both hardware performance metrics and reconstruction quality into its reward function. Table I provides a comprehensive comparison between our work and previous approaches, highlighting the unique capabilities of HERO that are not supported by existing methods. 

To summarize, our work makes the following contributions:
\begin{itemize}
\item We introduce HERO, the \textit{first} automatic hardware-aware NeRF quantization framework that eliminates the need for domain expertise and manual intervention through RL-based optimization.
\item We enable adaptive bit width assignment across different hash table levels, allowing our framework to optimize each level independently for enhanced model compactness and computational efficiency beyond uniform quantization approaches.
\item We integrate a custom cycle-accurate hardware simulator that incorporates detailed memory architecture modeling to capture the impact of hash table bit widths on memory access patterns and their influence on overall latency.
\item Experimental results demonstrate that HERO achieves 1.33× and 1.31× lower latency, 1.33× and 1.29× higher cost efficiency, and significantly smaller model sizes compared to state-of-the-art methods in high-fidelity and resource-constrained scenarios, respectively.
\end{itemize}

\begin{figure}
    \centering
    \includegraphics[width=.8\linewidth]{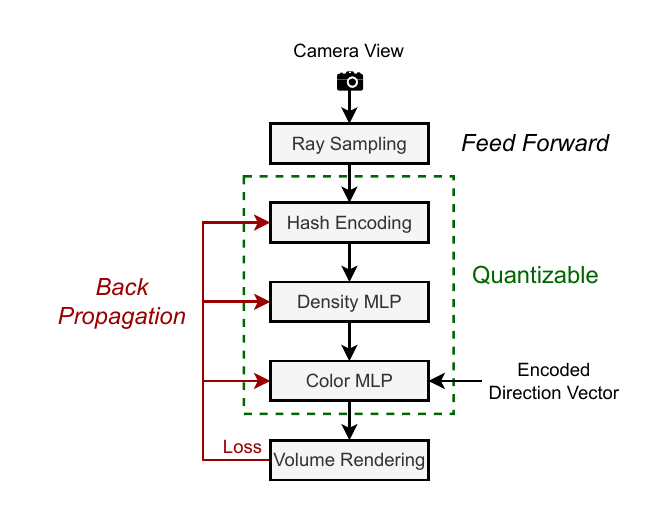}
    \caption{Instant NGP~\cite{muller2022instant} process}
    \label{fig:ngp}
    \vspace{-12pt}
\end{figure}

\section{Preliminary}

\subsection{Neural Radiance Field}

Vanilla NeRF~\cite{mildenhall2020nerf} has transformed the field of novel view synthesis through its innovative fusion of computer graphics methodologies with deep learning approaches. This groundbreaking contribution pioneers the application of deep neural networks for representing and rendering three-dimensional scenes using only a limited collection of input viewpoints. Nevertheless, the initial NeRF framework depends on computationally intensive MLPs characterized by extensive parameter counts. As a result, the rendering procedure becomes extremely time-intensive, frequently demanding hours or potentially days to produce a single synthesized view. 

To address this limitation, Instant NGP~\cite{muller2022instant} introduces significant advancements in both training and rendering efficiency. Fig.~\ref{fig:ngp} illustrates Instant NGP's training and rendering process, involving five main steps. It employs a multi-resolution hash table architecture coupled with two compact MLPs that replace the original large MLP. This hash encoding incorporates multiple levels of hash tables that store embedding vectors at different resolution levels. The key insight of the multi-resolution hash encoding is that it compresses the large voxel grid into multiple hash tables and eliminates unimportant information through potential hash collisions. However, the current Instant NGP model still contains substantial redundancy and can be further compressed through quantization techniques to enhance computational efficiency. The quantizable modules in Instant NGP, highlighted in green, present opportunities for quantization to reduce computational complexity.

\begin{figure}
    \centering
    \includegraphics[width=\linewidth]{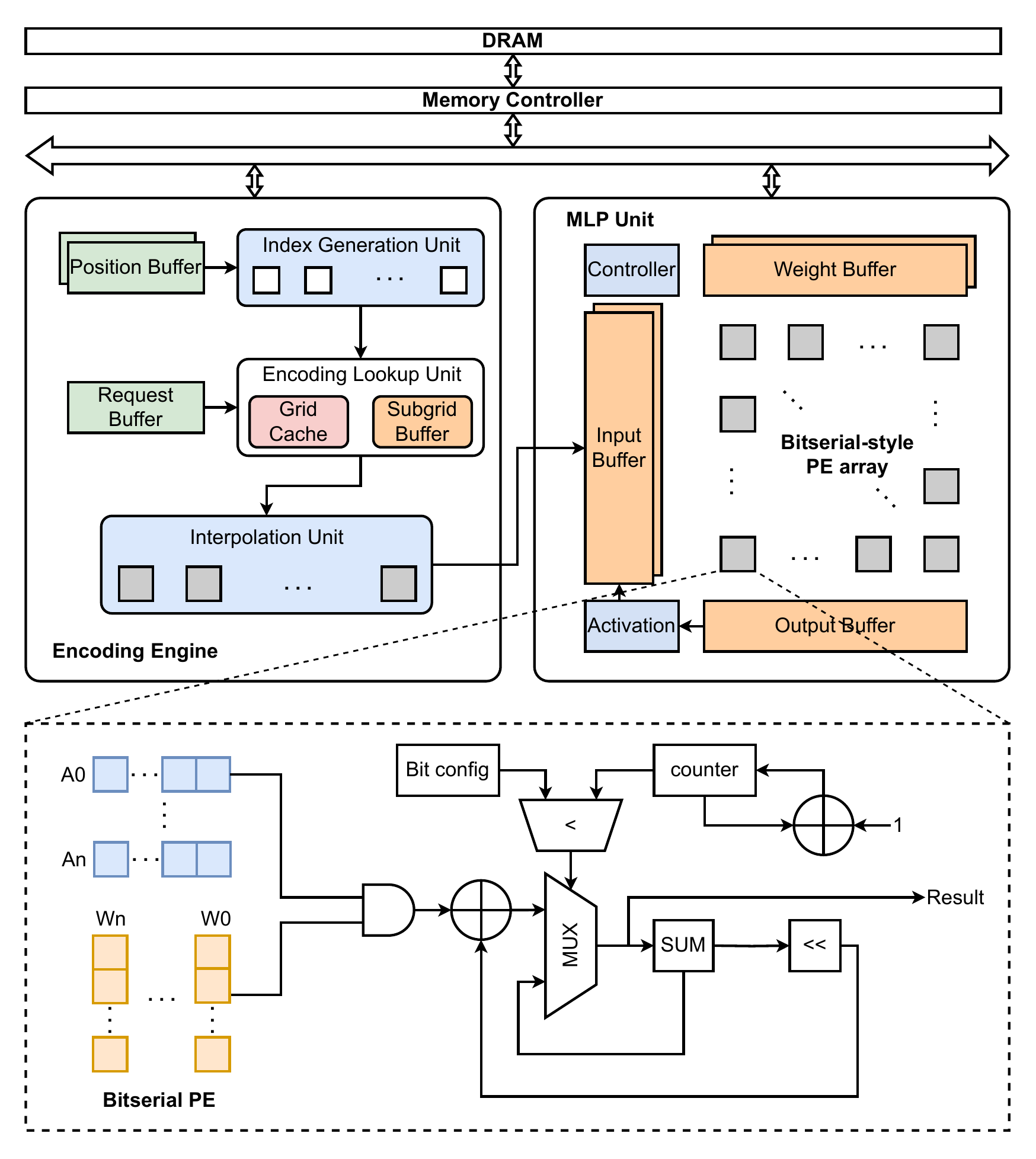}
    \caption{NeuRex-style~\cite{lee2023neurex} accelerator architecture with Bitserial PEs adopted to get hardware feedback in our work}
    \label{fig:neurex}
    \vspace{-12pt}
\end{figure}

\begin{figure*}
    \centering
    \includegraphics[width=\linewidth]{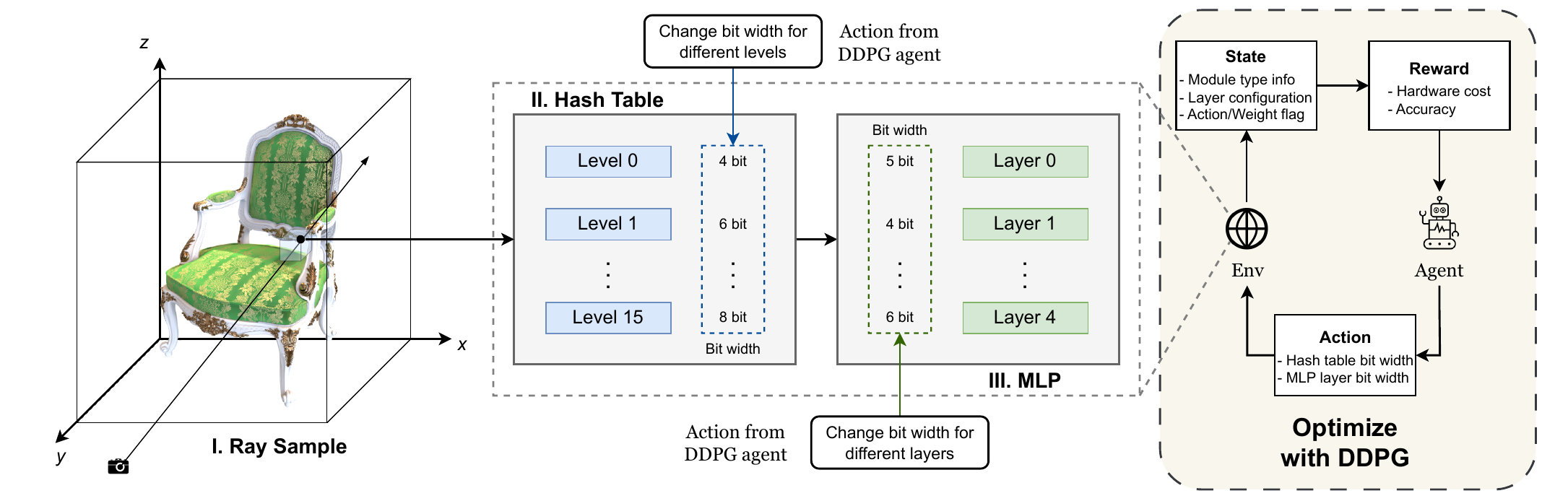}
    \caption{Overview for the proposed HERO framework based on DDPG agent}
    \label{fig:hero_flow}
    \vspace{-12pt}
\end{figure*}

\subsection{NeRF Accelerator Architecture}

Various hardware accelerators~\cite{li2023instant,lee2023neurex,ryu202420,zhang2025spnerf} have been developed to enhance NeRF training and rendering efficiency. Among these implementations, NeuRex~\cite{lee2023neurex} stands out due to its TPU-like systolic array architecture, which can potentially support mixed-precision computation by replacing processing elements (PEs) with BitSerial-style implementations, as shown in Fig.~\ref{fig:neurex}. NeuRex comprises two main components: an Encoding Engine and an MLP Unit. The Encoding Engine optimizes multi-resolution hash encoding through a hierarchical memory design. Coarse-level hash tables (levels 0-7) are stored in the grid cache to exploit higher data reuse from lower resolution and larger grid voxels, enabling multiple sample points to share voxel data. Fine-level hash tables (levels 8-15) reside in the subgrid buffer, where corresponding hash tables are preloaded on-chip for each subgrid. The heavily-banked subgrid buffer minimizes off-chip memory access by prefetching data only during subgrid transitions.

For our quantization implementation, we replace the original MLP Unit with BitSerial-style~\cite{judd2016stripes} processing elements that support arbitrary-precision integer multiplication. As shown in Fig.~\ref{fig:neurex}, the Bitserial PE architecture enables $N$-bit multiply-accumulate (MAC) operations to be computed in $N$ cycles, providing the flexibility needed for mixed-precision quantization strategies while maintaining computational efficiency.

\section{Methodology}

We propose HERO, a hardware-aware quantization framework that leverages a deep deterministic policy gradient (DDPG) agent to derive optimal quantization strategies by directly incorporating real-time latency feedback from our custom hardware simulator. Unlike traditional DNN quantization that targets only linear layers such as convolution and fully connected layers, our approach handles the unique architectural components of NeRF models. Specifically, hash table bit width selection impacts grid cache behavior in the underlying NeuRex architecture, affecting memory access patterns and overall system performance—a critical consideration that our custom simulator explicitly models to provide accurate hardware feedback. As illustrated in Fig.~\ref{fig:hero_flow}, HERO simultaneously optimizes quantization for both hash table encodings and MLP layers, dynamically determining their respective bit widths through actor-critic decision making. This comprehensive approach enables automatic hardware-aware quantization across the entire NeRF pipeline without requiring domain expertise or manual intervention, effectively exploring the extensive mixed-precision design space while maintaining the critical balance between reconstruction quality and computational efficiency.

\subsection{Observation Space}


To enable comprehensive quantization across both hash tables and MLP layers, we employ a unified seven-dimensional observation vector $S_i$ that treats each hash table level as equivalent to an MLP layer within the observation space. This design allows our DDPG agent to make consistent quantization decisions across heterogeneous NeRF components.

For MLP layers, the observation $S_i$ for the $i^{th}$ layer is defined as:
\begin{equation}
S_i = (L_i, d_{i}^{in}, d_i^{out}, W_i, i, a_{i-1}, f_{w/a}),
\end{equation}
where $L_i$ denotes the layer type indicator, $d_{i}^{in}$ and $d_{i}^{out}$ represent input and output dimensions respectively, $W_i$ indicates the weight parameter size, $a_{i-1}$ captures the previous quantization action, and $f_{w/a}$ distinguishes between activation ($f_{w/a}=0$) and weight ($f_{w/a}=1$) quantization.

For each level hash table, the observation is reformulated to capture hash-specific characteristics:
\begin{equation}
S_i = (L_i, d_i^{emb}, d_i^{en}, l_i, i, a_{i-1}, 1),
\end{equation}
where $d_i^{emb}$ represents the encoded embedding dimension, $d_i^{en}$ denotes the number of hash entries, $l_i$ indicates the specific hash table level, and $f_{w/a}=1$ since hash tables store feature parameters analogous to weights. This unified representation enables the agent to optimize bit width allocation across the entire NeRF architecture while accounting for the distinct computational characteristics of each component type.

\subsection{Action Space}

In the HERO framework, each action $a_i$ is represented as a continuous value within $[0,1]$ to enable smooth gradient flow during RL training. The corresponding quantization bit width for the $i^{th}$ layer is determined through:
\begin{equation}
b_i=\mathrm{round}(b_{min}-0.5+a_i\times[(b_{max}+0.5)-(b_{min}-0.5)])
\end{equation}
where $b_{min}=1$ and $b_{max}=8$, providing a comprehensive range from aggressive to conservative quantization. This continuous-to-discrete mapping maintains differentiability essential for policy gradient optimization while ensuring precise bit width control. The continuous representation preserves the relative ordering of quantization aggressiveness~\cite{wang2019haq}, allowing the agent to learn smooth transitions between different precision levels and effectively explore the quantization strategy space.

\subsection{Quantization}

Linear quantization is applied to both activations and weights to enable mixed-precision computation. Since our Bitserial-style processing elements support arbitrary-precision integer multiplication, we quantize values to integers to accurately simulate quantization-induced errors. 

For weights, we implement symmetric quantization centered at zero with the scale factor:
\begin{equation}
s = \frac{r_v}{2^{b_i} - 1},
\end{equation}
where $r_v = v_{max} - v_{min}$ represents the value range determined through calibration. The quantized weight value $q$ for input $x$ is:
\begin{equation}
q = \mathrm{clip}[\mathrm{round}(\frac{x}{s}),q_{min}, q_{max}],
\end{equation}
with quantization boundaries $q_{min} = -2^{b_{i}-1}-1$ and $q_{max} = 2^{b_{i}-1}-1$.

For activations, we employ asymmetric quantization with a non-zero centroid to better accommodate the typically non-negative activation distributions in NeRF models. The zero-point $Z$ is calculated as:
\begin{equation}
Z = \mathrm{round}[(1 - \frac{v_{max}}{r_v})\times(2^{b_i} - 1)].
\end{equation}
The quantized activation value $q$ for input $x$ is determined as:
\begin{equation}
q = \mathrm{clip}[\mathrm{round}(\frac{x}{s}+Z),0,2^{b_i}-1].
\end{equation}

\subsection{Reward}

To incorporate hardware efficiency into our optimization framework, we design a reward function $\mathcal{R}$ that balances reconstruction quality with computational cost:
\begin{equation}
\mathcal{R} = \lambda(PSNR_{cur} - PSNR_{org} + \frac{1}{cost\ ratio}),
\end{equation}
where the cost ratio is defined as:
\begin{equation}
cost\ ratio = \frac{current\ cost}{original\ cost}.
\end{equation}
Here, $original\ cost$ and $PSNR_{org}$ represent the baseline hardware latency and reconstruction quality obtained with all layers configured to maximum 8-bit precision, providing a reference point for evaluating quantization benefits. The $current\ cost$ reflects the actual hardware latency measured by our custom simulator for the current quantization configuration, while $PSNR_{cur}$ captures the corresponding reconstruction quality.

The scaling factor $\lambda = 0.1$ serves as a normalization factor to scale the reward magnitude appropriately for the RL training process. The reciprocal cost ratio term incentivizes the agent to minimize hardware latency—as computational cost decreases, the reward increases proportionally. This formulation enables our RL agent to discover quantization strategies that achieve optimal trade-offs between visual fidelity and hardware performance across diverse deployment scenarios.

\begin{table*}[tb]
  \caption{
  \textbf{Quantitative comparisons.} 
  Instant-NGP quantized with PTQ, QAT, and CAQ are compared.
  The latency ($\times 10^7 $ cycles per camera ray) and PSNR (dB) are reported to measure the performance and accuracy, respectively.
  }
  \label{tab:comparsion}
  \centering
  \setlength{\tabcolsep}{0.9mm} 
  \begin{tabular*}{\linewidth}{@{\extracolsep{\fill}}cp{2cm}cccccccc@{}}
  
    \toprule
    \multicolumn{2}{c}{\multirow{2}{*}{\centering Method} } 
    & \multicolumn{2}{c}{Chair}
    
    & \multicolumn{2}{c}{Lego}  
    & \multicolumn{2}{c}{Ficus}
    & \multicolumn{2}{c}{Average}
    \\ 
    \cmidrule[0.5pt](lr){3-4} \cmidrule[0.5pt](lr){5-6} \cmidrule[0.5pt](lr){7-8} \cmidrule[0.5pt](lr){9-10} 
    \multicolumn{2}{c}{\ }
    & \centering $\text{Latency}_\downarrow$ 
    & \centering $\text{PSNR}_\uparrow$ 
    & \centering $\text{Latency}_\downarrow$ 
    & \centering $\text{PSNR}_\uparrow$ 
    & \centering $\text{Latency}_\downarrow$ 
    & \centering $\text{PSNR}_\uparrow$
    & \centering $\text{Latency}_\downarrow$ 
    & {\centering $\text{PSNR}_\uparrow$ }
    \\
    \midrule
    & NGP       & / & 34.55 & / & 35.32 & / & 31.77 & / & 33.88
    \\
    \midrule
    \multirow{4}{*}{MDL} 
    & NGP-PTQ    & 4.84 & 26.07  & 0.73 & 26.01 & 2.10 & 27.05 & 2.56 & 26.38
    \\
    & NGP-QAT  & 4.84 & 31.27  & 0.73 & 32.30 & 2.10 & 28.82 & 2.56 & 30.80
    \\
    & NGP-CAQ& 6.32 & 34.07  & 1.02 & 34.76 & 2.57 & 31.41 & 3.30 & 33.41
    \\
    & \textbf{HERO (Ours)} & 4.70 & 33.16 & 0.71 & 33.29 & 2.04 & 29.59 & 2.48 & 32.01
    \\
    \midrule
    \multirow{4}{*}{MGL\ ($ 10^{-3.2}$)}
    & NGP-PTQ   & 4.03 & 23.93 & 0.61 & 25.34 & 1.75 & 26.04 & 2.13 & 25.10
    \\
    & NGP-QAT  & 4.03 & 30.78 & 0.61 & 30.79 & 1.75 & 27.78 & 2.13 & 29.78
    \\
    & NGP-CAQ & 5.33 & 31.34 & 0.75 & 30.96 & 2.33 & 29.66 & 2.80 & 30.65
    \\
    & \textbf{HERO (Ours)} & 3.88 & 29.90 & 0.58 & 31.72 & 1.92 & 28.13 & 2.13 & 29.92
    \\
    \bottomrule
  \end{tabular*}
  \vspace{-12pt}
\end{table*}

\subsection{Agent and Environment}

We employ the DDPG algorithm—an off-policy actor-critic method designed for continuous action spaces—to optimize our quantization strategy. Within our framework, the DDPG agent operates episodically, sequentially determining the bit width for each layer across the entire NeRF architecture. Upon completing the bit width assignment for all layers, we perform model retraining to restore reconstruction quality, followed by PSNR evaluation to compute the reward signal that guides policy optimization.

The expected return Q-function in our framework incorporates variance reduction through:
\begin{equation}
\hat{Q_i} = \mathcal{R} + \gamma \times Q[S_{i+1}, \mu(S_{i+1})|\theta^Q] - \epsilon,
\end{equation}
where $\epsilon$ represents the exponential moving average of previous rewards to mitigate variance in gradient estimation, $\mu$ denotes the deterministic policy function (actor), and $\gamma$ is the discount factor. This formulation stabilizes training by reducing the impact of reward fluctuations inherent in the hardware simulation and reconstruction quality evaluation process.

The corresponding critic loss function is computed as:
\begin{equation}
\mathcal{L} = \frac{1}{K_a}\sum_{n = 1}^{K_a}[\hat{Q_i} - Q(S_i,a_i|\theta^Q)]^2,
\end{equation}
where $K_a$ represents the total number of quantization decisions (actions) executed within a single episode, corresponding to the sum of hash table levels and MLP layers in the NeRF architecture. This episodic approach enables the agent to learn dependencies between quantization choices across different network components while optimizing the overall hardware-quality trade-off.

\subsection{Hardware simulator}
To accurately assess hardware performance, we develop a cycle-accurate simulator that models the execution characteristics of a NeuRex-style accelerator. The simulation trace files are generated by executing real-world datasets on GPU platforms. We adopt identical timing and memory configurations as in~\cite{lee2023neurex}, specifically employing a 1 GHz clock frequency and LPDDR4-3200 memory configuration. We implement the same direct-mapped cache configuration for grid cache in NeuRex to accurately capture the impact of hash table bit widths on memory access patterns and cache behavior. Additionally, we modify the processing elements within the systolic array architecture, adopting Bitserial-style PEs to support arbitrary-precision integer MAC operations, thereby enabling flexible quantization bit width.

\section{Evaluation}

\subsection{Experiment Setup}
\textbf{Dataset and Model.}  Our proposed HERO framework is based on Instant NGP~\cite{muller2022instant}, a widely recognized NeRF model known for its exceptional computational efficiency. We implement our framework by extending the open-source codebase used in CAQ~\cite{liu2024content} and evaluate reconstruction quality using the Synthetic-NeRF dataset~\cite{mildenhall2020nerf}, which serves as a common benchmark in neural radiance field research.



\textbf{Baseline.} For our comparative experiments, we select CAQ~\cite{liu2024content}, a state-of-the-art NeRF quantization framework, as our baseline. CAQ introduces a novel approach to identify scene-dependent bitwidths that effectively reduce computational complexity in NeRF. The authors present experiments across distinct operational levels: MDL, which addresses high-fidelity rendering scenarios, and MGL, which targets resource-constrained environments. Within the MGL framework, a configurable target loss metric governs the target computational complexity, where larger values correspond to more aggressive optimization (lower complexity). Here we set the target loss to $10^{-3.2}$ to make it consistent with CAQ for comparison, represented as MGL($10^{-3.2}$). We further enhance our comparative analysis by including both Post-Training Quantization (PTQ) and Quantization-Aware Training (QAT)~\cite{bhalgat2020lsq+, ye2023mcunerf} approaches as the baseline. PTQ represents a direct approach, where predetermined bit widths are applied to the pre-trained model without additional optimization. In contrast, QAT incorporates quantization effects during the training process, fine-tuning the model parameters after initial quantization to recover accuracy losses. Following the methodology established in \cite{liu2024content}, we apply uniform quantization across all MLP layers—specifically 6-bit precision for MDL and 5-bit precision for MGL. These established quantization methods provide a comprehensive baseline against which we evaluate our proposed technique across various computational efficiency targets.

\begin{figure*}
     \centering
    
    \begin{minipage}[b]{0.48\textwidth} 
        \centering
        \includegraphics[width=.8\textwidth]{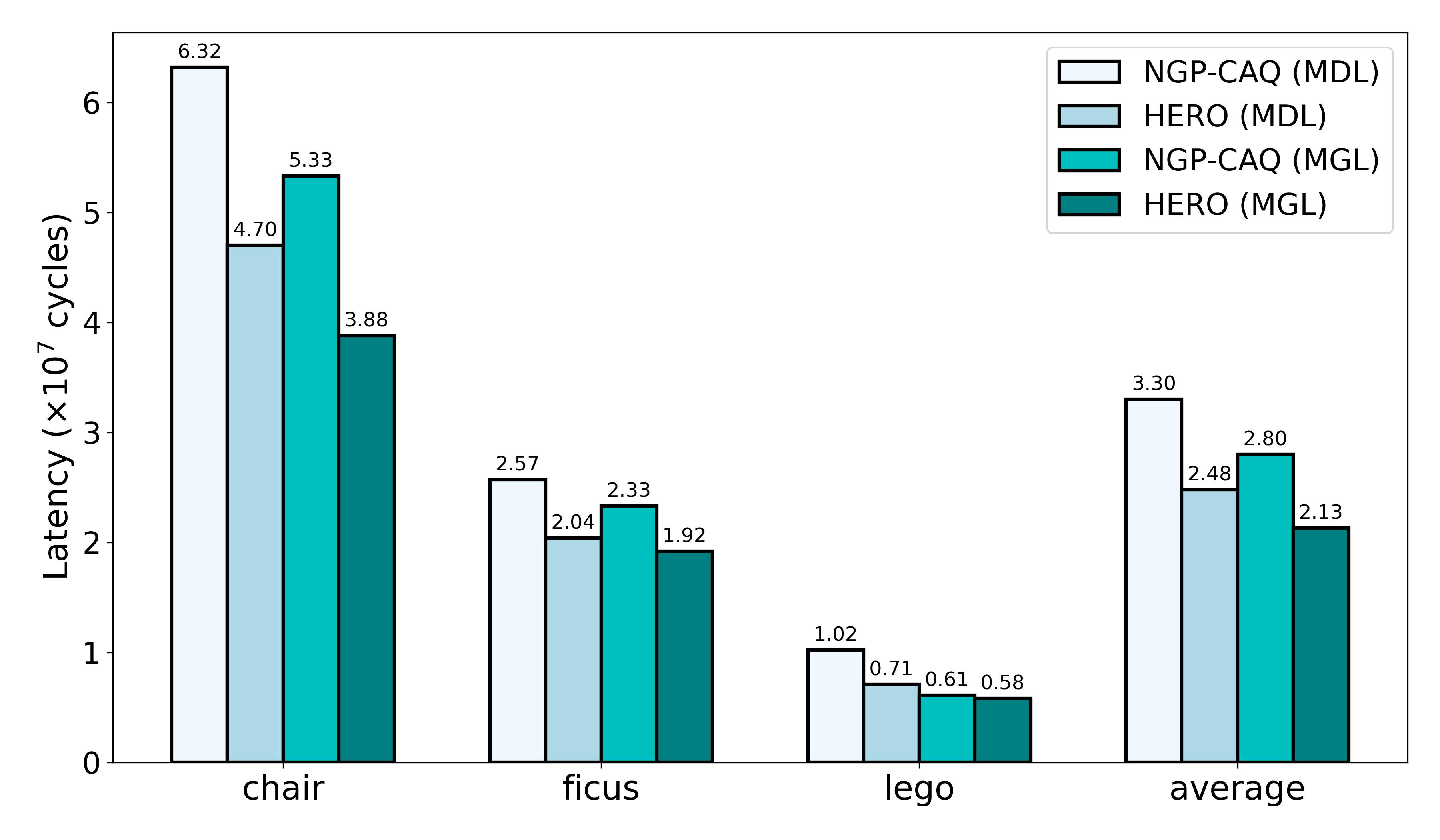}
        \subcaption{Latency comparison: NGP-CAQ vs. HERO}
        \label{fig:latency}
    \end{minipage}
    \begin{minipage}[b]{0.48\textwidth} 
        \centering
        \includegraphics[width=.8\textwidth]{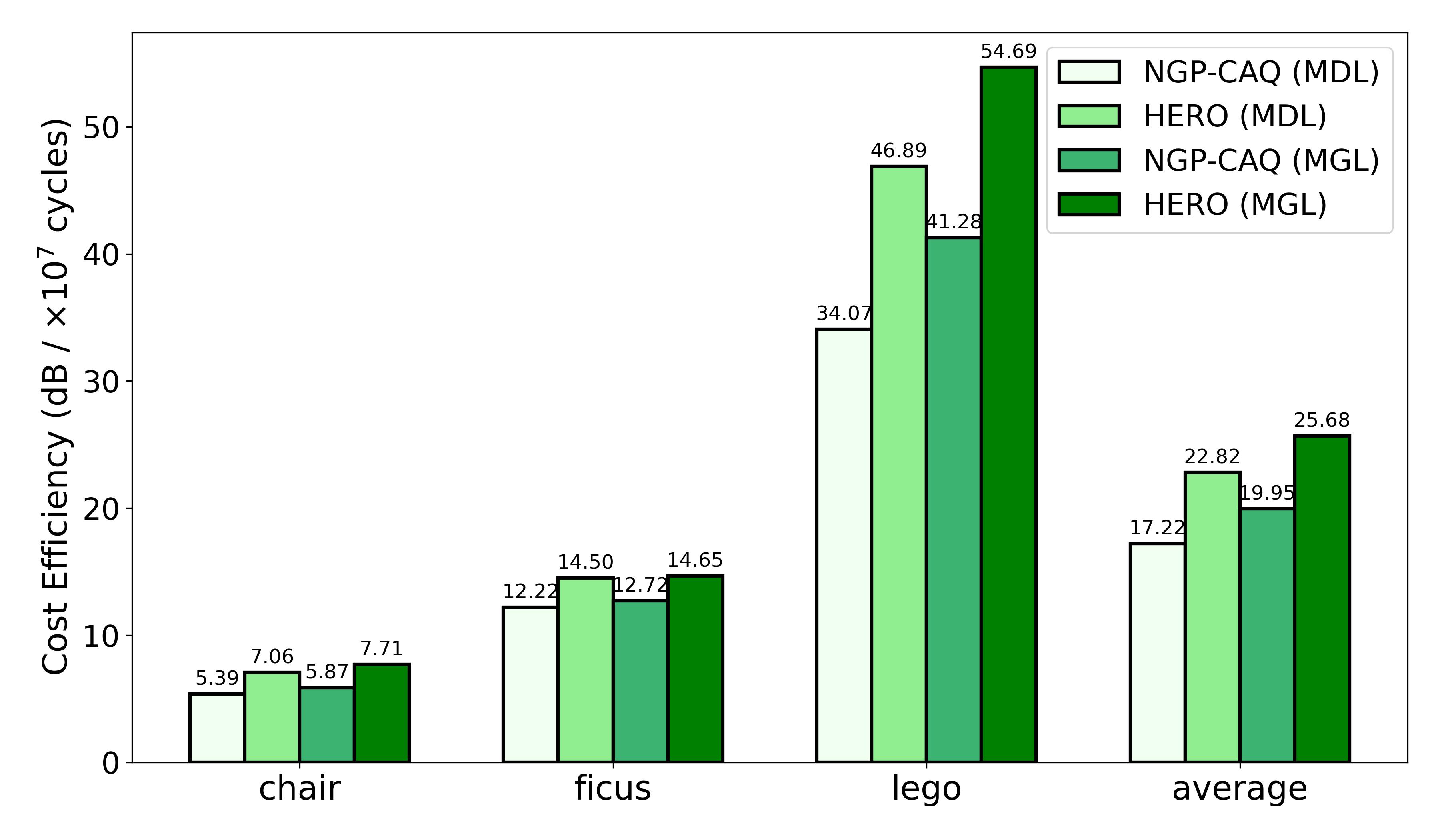}
        \subcaption{Cost Efficiency: NGP-CAQ vs. HERO}
        \label{fig:cost}
    \end{minipage}
    \caption{Latency and cost comparison in different levels and datasets}
    \label{fig:result}
    \vspace{-12pt}
\end{figure*}

\subsection{Evaluation Metrics}
To comprehensively assess the performance of our proposed framework, we employ multiple evaluation metrics that capture different aspects of model efficiency and quality.

\begin{table}[tb]
  \caption{
  \textbf{Model size comparisons.} 
  The FQR results of different quantization methods are compared to evaluate the model size.
  }
  \label{tab:fqr}
  \centering
  \setlength{\tabcolsep}{0.9mm} 
  \begin{tabular*}{\linewidth}{@{\extracolsep{\fill}}cp{2cm}cccc}
    \toprule
    \multicolumn{2}{c}{Method} 
    & Chair
    & Lego  
    & Ficus
    & Average \\
    \midrule
    & NGP       & 32.00 & 32.00 & 32.00 & 32.00 \\
    \midrule
    \multirow{4}{*}{MDL} 
    & NGP-PTQ    & 7.60 & 7.60  & 7.60  & 7.60 \\
    & NGP-QAT    & 7.60 & 7.60  & 7.60  & 7.60 \\
    & NGP-CAQ    & 9.16 & 9.33  & 9.67  & 9.39 \\
    & \textbf{HERO (Ours)} & 6.33 & 6.33  & 6.17  & 6.28 \\
    \midrule
    \multirow{4}{*}{MGL\ ($10^{-3.2}$)}
    & NGP-PTQ    & 6.60 & 6.60 & 6.60 & 6.60 \\
    & NGP-QAT    & 6.60 & 6.60 & 6.60 & 6.60 \\
    & NGP-CAQ    & 7.50 & 6.83 & 8.17 & 7.50 \\
    & \textbf{HERO (Ours)} & 5.17 & 5.17 & 6.00 & 5.45 \\
    \bottomrule
  \end{tabular*}
  \vspace{-12pt}
\end{table}

\textbf{Reconstruction Quality and Hardware Feedback.} We evaluate reconstruction quality using PSNR score and measure hardware performance through inference latency on the target accelerator platform. These metrics provide direct feedback for our RL agent to optimize the trade-off between reconstruction fidelity and computational efficiency.

\textbf{Cost Efficiency.} To evaluate the balance between computational efficiency and output quality, we introduce a cost efficiency metric that quantifies the relationship between hardware performance and reconstruction quality:

\begin{equation}
\text{Cost Efficiency} = \frac{\text{PSNR}}{\text{Latency}}.
\end{equation}

This metric provides a straightforward assessment of how effectively each approach converts computational resources into visual quality, with higher values indicating more efficient utilization of hardware capabilities. The cost efficiency metric enables direct comparison between different quantization strategies by normalizing quality gains against their computational overhead.

\textbf{Feature Quantization Rate (FQR).} We adopt the feature quantization rate from~\cite{liu2024content} to evaluate the quantized model size and compression effectiveness:

\begin{equation}
FQR = \frac{\sum_{i\in M}b_i}{M},
\end{equation}
where $b_i$ represents the bit width for the $i^{th}$ layer, and $M$ is the total number of layers in the model. A lower FQR indicates more aggressive quantization and smaller model size, while maintaining the balance between compression and accuracy.

\subsection{Quantitative Results}

Table \ref{tab:comparsion} presents our comprehensive comparative analysis. It is worth noting that PTQ and QAT exhibit identical latency profiles at each optimization level (MDL and MGL), as we intentionally maintain identical bit widths for both approaches to ensure fair comparison.  In contrast, CAQ employs a more sophisticated approach, dynamically allocating different bit widths to individual layers through a search-based optimization that targets specific performance criteria. However, our analysis reveals a notable limitation in CAQ's optimization strategy: it prioritizes reconstruction quality while inadequately addressing computational efficiency, resulting in suboptimal latency performance and a notably large model size.

\begin{figure}
    \centering
    \includegraphics[width=\linewidth]{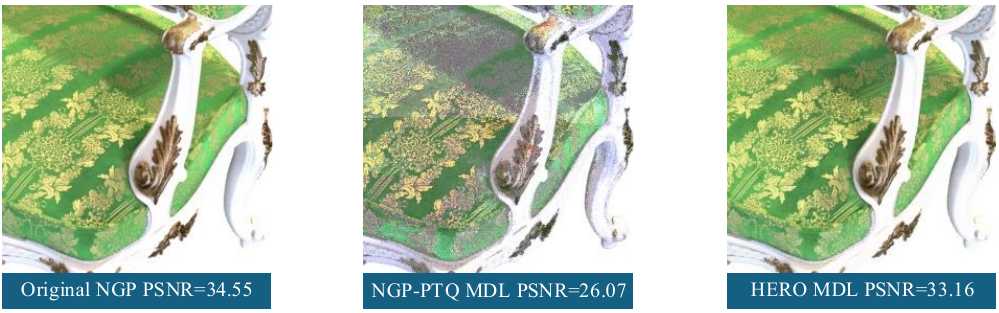}
    \caption{Qualitative results: Original NGP with full precision, NGP-PTQ in MDL level, and HERO (Ours) in MDL level}
    \label{fig:visual}
    \vspace{-12pt}
\end{figure}

The computational latency and cost efficiency results are presented in Fig. \ref{fig:result}. While CAQ achieves higher PSNR values, it suffers from notable computational overhead, exhibiting 1.33× and 1.31× higher latency at MDL and MGL levels, respectively, along with substantially larger model sizes as shown in Table~\ref{tab:fqr}. This performance degradation stems from CAQ's optimization strategy, which prioritizes PSNR preservation without considering latency. CAQ typically quantizes either activations or weights to lower bit widths while maintaining the other at higher precision, creating computational imbalances that increase processing overhead.

In contrast, HERO achieves 1.33× and 1.29× higher cost efficiency at MDL and MGL levels through hardware-aware optimization that incorporates direct latency feedback from our custom simulator. Our framework dynamically adjusts bit width configurations when performance metrics exceed predefined latency targets, ensuring balanced quantization across the entire pipeline. Additionally, while CAQ applies uniform bit widths across all hash table levels, HERO assigns different bit widths to different hash table levels, which reduces the overall FQR and results in smaller model sizes without sacrificing reconstruction quality.
This comprehensive approach enables HERO to deliver superior reconstruction quality per unit of computational cost while maintaining compact model representations, effectively optimizing the critical trade-off between visual fidelity and hardware efficiency that is essential for practical NeRF deployment.

\subsection{Qualitative Results}

Fig.~\ref{fig:visual} presents qualitative comparisons between different quantization methods using magnified portions from the chair dataset to highlight rendering detail preservation. The visual results demonstrate that HERO maintains high rendering quality without introducing noticeable artifacts, achieving comparable visual fidelity to the original full-precision Instant NGP implementation while delivering substantially reduced computational latency. In contrast to standard PTQ approaches that often exhibit degraded detail reconstruction, our hardware-aware quantization strategy preserves fine-grained features and texture details, validating the effectiveness of our RL-guided bit width allocation across hash table levels and MLP layers.

\section{Conclusion}

This paper introduces HERO, the \textit{first} fully automated framework for NeRF quantization that integrates hardware feedback without requiring human expert intervention. Building upon Instant NGP, our approach achieves reconstruction quality comparable to full-precision models while effectively reducing hardware latency for NeuRex-style accelerators. Experimental results demonstrate that our framework achieves up to 1.33× lower latency, 1.33× higher cost efficiency, and smaller model sizes compared to current state-of-the-art methods. By using RL to search optimal bit width allocation across hash table levels and MLP layers with feedback from our custom cycle-accurate simulator, our framework effectively balances visual fidelity with computational demands. These contributions make neural radiance fields more practical for resource-constrained deployment scenarios while maintaining high-quality rendering performance.

\section*{Acknowledgment}
This work was partially supported by AI Chip Center for Emerging Smart Systems (ACCESS), Hong Kong SAR and Collaborative Research Fund (UGC CRF) C5032-23G.

\AtNextBibliography{\small}
\printbibliography{}

\end{document}